# Scalable $T^2$ resistivity in a small single-component Fermi surface

Xiao Lin[1], Benoît Fauqué[1], and Kamran Behnia[1]*

[1]Laboratoire de Physique et Etude des Matériaux (CNRS/UPMC), ESPCI, 10 Rue Vauquelin, F-75005, Paris, France

*Correspondence to: kamran.behnia@espci.fr.

**Abstract**: Scattering among electrons generates a distinct contribution to electrical resistivity that follows a quadratic temperature dependence. In strongly-correlated electron systems, the prefactor A of this $T^2$ resistivity scales with the magnitude of the electronic specific heat, γ. Here, we show that one can change the magnitude of *A* by four orders of magnitude in metallic $SrTiO_3$ by tuning the concentration of the carriers and consequently, the Fermi energy. The $T^2$ behavior persists in the single-band dilute limit despite the absence of two known mechanisms for $T^2$ behavior, distinct electron reservoirs and Umklapp processes. The results highlight the absence of a microscopic theory for momentum decay through electron-electron scattering in different Fermi liquids.

**Main Text:** Warming a metal enhances its resistivity because with increasing temperature (T) scattering events along the trajectory of a charge-carrying electron become more frequent. In most simple metals the dominant mechanism is scattering by phonons leading to a $T^5$ dependence of resistivity. In 1937, Baber identified electron-electron scattering as the origin of $T^2$ resistivity observed in many transition metals (1). During the last few decades it has been firmly established that, at low temperatures, resistivity (ρ) in a Fermi liquid follows a quadratic temperature dependence expressed as $\rho=\rho_0+AT^2$ and that correlations among electrons enhance both $A$ and the electronic specific heat, $\gamma$. This is often expressed through the Kadowaki-Woods ratio (2-6), $R_{KW}=A/\gamma^2$, which link two distinct properties of a Fermi liquid, each set by the same material-dependent Fermi energy, $E_F$.

The Pauli exclusion principle is the ultimate reason behind both the T-linear specific heat and T-square resistivity in Fermi liquids. Electrons that give rise to both properties are those confined to a width of $k_BT/E_F$, where $k_B$ is the Boltzmann constant. In the case of resistivity, this is true of both electrons participating in the scattering event, hence the exponent of two. However, electron-electron scattering alone does not generate a finite contribution to resistivity, because such a scattering event would conserve momentum with no decay in the charge current. The presence of an underlying lattice is required in any scenario for generating $T^2$ resistivity from electron-electron scattering. Dimensional considerations imply:

$$A = \frac{\hbar}{e^2} \left(\frac{k_B}{E_F}\right)^2 \ell_{quad} \quad (1)$$

Here, $\hbar$ and $e$ are fundamental constants and $l_{quad}$ is a material-dependent length scale, which can be set either by the Fermi wave-length of electrons, or by the interatomic distance or a combination of both. Mott argued that the average distance between two scattering events is proportional to the concentration and the collision cross section of electrons $\sigma_{cs}$ (7). Therefore:

$$A = \frac{\hbar}{e^2} \left(\frac{k_B}{E_F}\right)^2 k_F \, \sigma_{cs} \quad (2)$$

Here, $k_F$ is the Fermi wave-vector and $\sigma_{cs}$ is set by the specific process governing the decay in charge current due to the presence of lattice.

There are several types of theoretical proposals for generating $T^2$ resistivity from electron-electron scattering in the presence of a lattice. The first (1) invokes a multi-band system with two different electron masses. Momentum transfer between these two distinct electron reservoirs sets the temperature dependence of resistivity and the mass mismatch leads to a leak of momentum towards the lattice thermal bath. The second invokes Umklapp scattering and the fact that momentum conservation does not prohibit transferring a unit vector of the reciprocal lattice (8, 9). In addition to these, it has been recently argued that Fermi liquids lacking Galilean invariance, which have non-parabolic and anisotropic energy dispersions, can display T-square resistivity even in the absence of any Umklapp process (10) thanks to electron-impurity scattering. In addition to these semi-classic scenarios, quantum interference can also generate a resistivity proportional to $T^2 \ln T$ (10, 11). The relevance of these ideas to the ubiquitous $T^2$ resistivity observed in a wide variety of Fermi liquids has not been settled experimentally.

It has been known for two decades that n-doped $SrTiO_3$ with a carrier density exceeding 0.01 e$^-$ per formula unit (f.u.) follows a $T^2$ resistivity (12). This $T^2$ resistivity provided input for the analysis of Kadowaki-Woods ratio in low-density Fermi liquids (5) and the Landau quasi-particles of the polaron Fermi liquid (13). More recently, it has been reported that due to its exceptionally long Bohr radius, $SrTiO_3$ keeps a robust metallic resistivity down to very low doping levels (14). Moreover, both oxygen-deficient (15, 16) and La-doped $SrTiO_3$ (17) host a well-defined Fermi surface down to carrier densities as low as 3 10$^{17}$ cm$^{-3}$ (which corresponds to 2 10$^{-5}$e$^-$ per f.u.). Such a context provides a unique opportunity to test the relevance of different theoretical pictures for the origin of $T^2$ resistivity.

Here we present resistivity measurements that show that the $T^2$ resistivity persists when carrier density becomes two orders of magnitude lower than what was reported before (12,13) . The magnitude of $A$ varies smoothly as a function of $E_F$ and becomes comparable to what has been seen in a heavy-fermion metal. The most important finding is the persistence of $T^2$ behavior in the single-band regime, where there is only a single electron reservoir with a Fermi wave-vector much too small for any Umklapp process. This severely restrains possible origins of the observed $T^2$ resistivity. The experimental determination of the collision cross section of electrons in a Fermi liquid with a simple and well-documented Fermi surface topology provides a quantitative challenge for theory. Comparing the data obtained on n-doped $SrTiO_3$ with other Fermi liquids, we argue that $l_{quad}$, the characteristic length scale of e-e scattering in each Fermi liquid, is a source of information regarding the microscopic origin of momentum decay.

The evolution of resistivity as carrier density changes from between $10^{17}$ and $10^{20}$ cm$^{-3}$ is presented in Fig.1 (see [18] for details on all 35 samples studied). In agreement with previous reports (14-17, 19), $SrTiO_3$ in this doping range is found to be a dilute metal whose resistivity drops by several orders of magnitude as it is cooled down from room temperature to liquid helium temperatures.

Above 100 K, the scattering rate extracted from resistivity and carrier concentration ($\tau^{-1} = \frac{\rho n e^2}{m_e}$ in Fig. 1B) does not vary with doping and follows roughly a $T^3$ dependence (we are neglecting the mass renormalization, which would lead to a correction between 1.8 and 5 in this doping window). Below 100 K, inelastic resistivity evolves with carrier concentration. Both electron-phonon and electron-electron scattering mechanisms can depend on the size of the Fermi surface. In the case of acoustic phonons, as documented in graphene (20), the Bloch-Grüneisen temperature ($\Theta_{BG} = \frac{2\hbar v_s k_F}{k_B}$ where $v_s$ is the sound velocity) separates two regimes. In a degenerate three-dimensional Fermi liquid, the inelastic resistivity caused by phonon scattering is expected to follow $T^5$ below $\Theta_{BG}$ and become T-linear above $\Theta_{BG}$. In our case, $\Theta_{BG}$ and the Fermi degeneracy temperature are of the same order of magnitude. Therefore, at high temperatures, electrons scattered by phonons are obeying Boltzmann statistics. Here, we focus on the $T^2$ inelastic resistivity emerging at low temperatures, which has been attributed to the scattering of electrons off each other (5, 12, 13).

As seen in Fig. 1, D-F, the slope of $\rho$ vs. $T^2$ plots in SrTiO$_{3-\delta}$ smoothly decreases with increasing carrier concentration. In all cases, there is a deviation upward from the $T^2$ behavior towards a regime with a higher exponent. This is to be contrasted with the case of Fermi liquids with strong correlation, in which quasi-particles are destroyed by warming well below the degeneracy temperature. In SrTiO$_3$, the temperature at which the deviation occurs increases with doping. We found similar behavior in Nb-doped and La-doped SrTiO$_3$ (18).

Figure 2A shows that the magnitude of $A$ as a function of carrier concentration. Our data is compatible with what has been previously reported for higher carrier concentrations (12, 13). Thus, decreasing carrier concentration is concomitant with a monotonous and uninterrupted increase in the magnitude of $A$ across several orders of magnitude as expected from Equation 1. The residual resistivity, $\rho_0$, (inset) varies much less with carrier concentration. Figure 1C shows that the magnitude of $A$ in two samples with identical carrier densities, but different residual resistivities, is quasi-identical. Therefore the magnitude of $A$ is set by $n$ and not by $\rho_0$.

In a Fermi liquid, the Fermi energy is reduced either when the Fermi surface shrinks or when the effective mass is enhanced. In both cases, the magnitude of $A$ is expected to enhance according to Equation 1. Mass enhancement is the origin of the large $A$ in heavy-fermion metals. Our results show that a large $A$ can also be achieved by reducing the sheer size of the Fermi surface.

In the extreme dilute limit, *A* becomes an order of magnitude larger than what is found in heavy-fermion UPt$_3$ (22).

Figure 2A reveals a hump in *A(n)* near $n=1.2 \times 10^{18}$cm$^{-3}$. According to an extensive study of quantum oscillations (16), at this carrier density, dubbed $n_{c1}$, a second band begins to be filled and the cyclotron mass of the lowest band suddenly enhances. Figure 2B shows the energy dispersion in the two bands constructed from the frequency and effective mass obtained by quantum oscillations (18). The deviation from parabolicity in the lowest band occurs at k = 0.4 nm$^{-1}$, close to the expectations from the theoretical band structure, according to which anti-crossing between bands generates a downward deviation of the lowest band near this wave-vector (13).

The dispersion map of Figure 2B allows us to determine the Fermi energy of each sample from its carrier density, leading to Fig. 2C, which shows *A* as a function of the Fermi energy of the lowest band with no visible anomaly near $n_{c1}$. The dependence remains close to $E_F^{-2}$ over a wide range. This is a strong indication that the $n_{c1}$ anomaly seen in Fig. 1A is almost entirely caused by deviation from parabolic dispersion in the lowest band, which hosts most of carriers.

As seen in Figure S2 (18), one can clearly detect a correlation between large *A* and small $E_F$ across different materials by comparing the variation of *A* with Fermi energy in SrTiO$_{3-\delta}$ and in other Fermi liquids. This is an extension of the Kadowaki-Woods approach to include dilute Fermi liquids in which the electronic specific heat, I set by the ratio of carrier density to the Fermi energy.

Using Eq. 1, one can extract $l_{quad}$, the characteristic length scale associated with electron-electron scattering in SrTiO$_{3-\delta}$. The extracted length (Figure 3A) shows only a very slight decrease with doping and is *not* proportional to the Fermi wave-length. Note that this would have led to an $n^{-5/3}$ dependence of *A* in conformity with the simplest available treatments of electron-electron scattering (22, 23).

Figure 3A compares the magnitude of $l_{quad}$ in $SrTiO_{3-\delta}$ with other Fermi liquids (See tables S3 and S4 for details). In a multi-component Fermi surface, a complication arises because there is a multiplicity of Fermi energies. When the Fermi surface occupies a large fraction of the Brillouin zone, one can assume that there is roughly one electron per f.u. and extract the Fermi energy from γ. This assumption allows one to extract an order of magnitude estimate for $l_{quad}$ in dense heavy-fermion and transition metals. In Fig. 3A, they lie close to the horizontal lines representing the Kadowaki-Woods ratios in the two families (2- 6). Figure 3A also includes data for the Fermi-liquid unconventional superconductor $Sr_2RuO_4$ (24), the heavily-doped non-superconducting LSCO (25), and the YBCO cuprate at p=0.11 (in which resistivity is $T^2$ (26, 27) and the Fermi energy of the small pocket seen by quantum oscillation has been quantified (28)). We have also included reported data on bismuth (29), on graphite (30) and on arsenic-doped germanium (31). Figure 3A shows that $l_{quad}$ lies mostly between 1 and 40 *nm*. Its magnitude is to be linked to the microscopic details of momentum decay by scattering in each system.

Using Eq. 2, we have also extracted the collision cross section of electrons in $SrTiO_{3-\delta}$ from the magnitude of *A* and the measured radius of the Fermi surface. Figure 3B shows its variation as a function of their Fermi wave-length. If the electrons were classical objects bouncing off each other, $\sigma_{cs}$ would have been $2\pi\lambda_F^2$. Our data are inconsistent with that classical picture; Figure 3B shows that $\sigma_{cs}$ is much smaller than $2\pi\lambda_F^2$ and does not follow $\lambda_F^2$. It remains a theoretical challenge to provide a quantitative explanation for this observation.

The principal conclusion of this study is that a comprehensive understanding of how $T^2$ resistivity is caused by electron-electron scattering in Fermi liquids is missing. Previous to this work, $T^2$ resistivity in Fermi liquids was observed in systems with a large single-component Fermi surface (such as $La_{1.7}Sr_{0.3}CuO_4$) or those with small multi-component ones (such as bismuth or graphite). In each case, one of the scenarios sketched in Figure 4 could be ruled out. However, one could still invoke either the multiplicity of reservoirs or the relevance of Umklapp processes. In the case of extremely dilute $SrTiO_{3-\delta}$ no room is left for either of the two. An Umklapp event can only occur if the largest available Fermi wave-vector is one-fourth of the smallest vector of the reciprocal lattice, $G$. By a rough estimation this corresponds to a carrier density of $2\ 10^{20}$ cm$^{-3}$ and Umklapp scattering may cause the hump in the energy dependence of $A(E_F)$ near 10meV (see Figure 2C) corresponding to this carrier concentration. However, we find that $A$ is still growing when $k_F$ becomes thirty times smaller than $G$.

In the specific case of doped $SrTiO_3$, an explanation of the $T^2$ resistivity may invoke the polaronic nature of the quasi-particles (13) or the distorted structure of the Fermi surface (17). Beyond this particular case, our results highlight the absence of a microscopic theory for momentum decay through electron-electron scattering in different Fermi liquids. Future experiments can quantify the magnitude of $l_{quad}$ in each of them. A possibly significant role of phonon-assisted (32) electron-electron scattering is to be reconsidered.

**Acknowledgments:** We thank H. Maebashi, D. Maslov and K. Miyake for stimulating discussions. This work was supported by Agence Nationale de Recherche through the SUPERFIELD project and by a grant attributed by the *Ile de France* region.

Supplementary Materials:

Materials and Methods

Figures S1 and S2

Table S1 to S4

References (33-60)

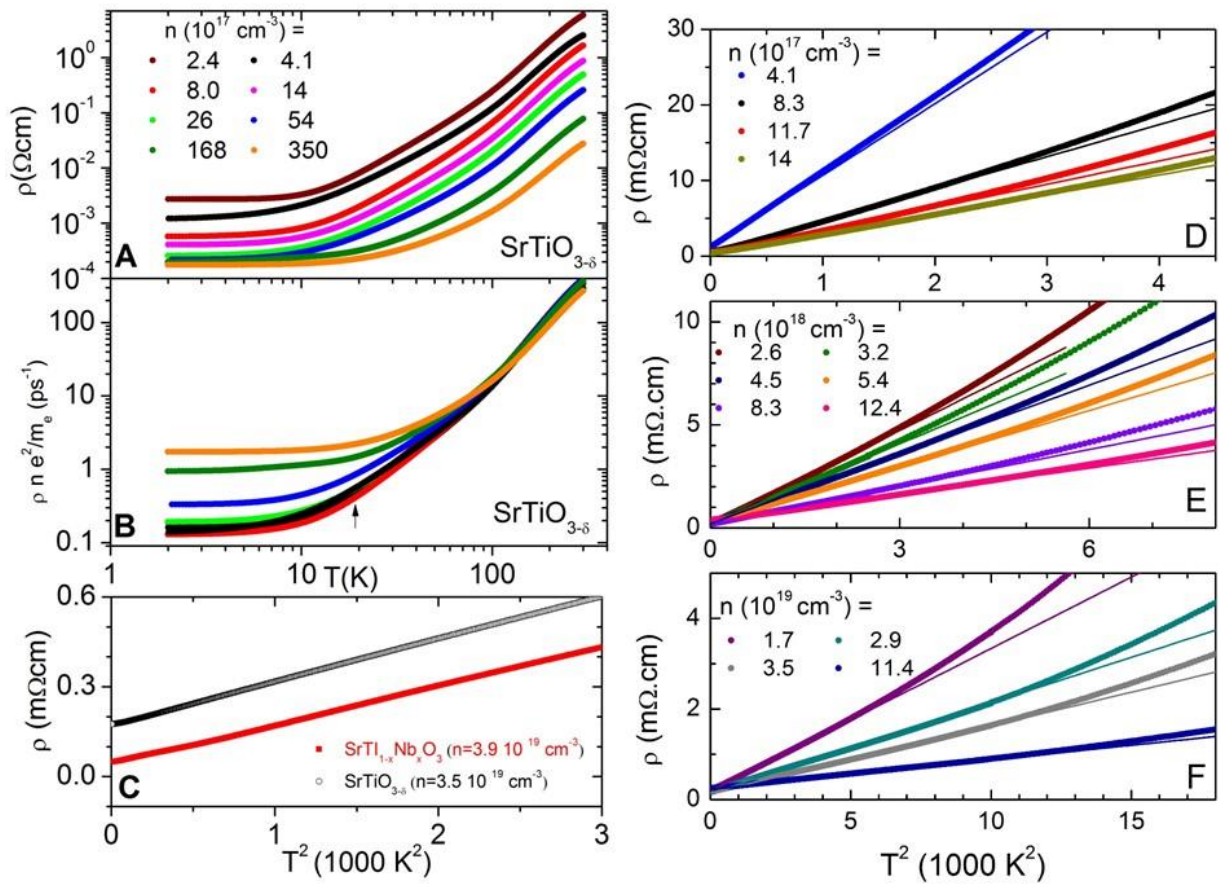

**Fig. 1**. Doping and temperature dependence of resistivity in n-doped $SrTiO_3$: (A) Evolution of resistivity in $SrTiO_{3-\delta}$ with doping across two orders of carrier density. (B). The product of resistivity and carrier density yields the scattering rate, which does not depend on carrier concentration above 100 K. (C) Resistivity plotted as a function of $T^2$ in oxygen-deficient and Nb-doped $SrTiO_3$ samples of comparable carrier concentration displays the same slope but different intercepts. (D –E) Resistivity vs. $T^2$ in $SrTiO_{3-\delta}$ as the carrier density changes by two orders of magnitude. Solid lines are straight lines representing the best fit to low-temperature data. As doping increases, the slope gradually decreases and the upward deviation towards the phonon-dominated regime shifts to higher temperatures. Note the change in the vertical and horizontal scales with increasing carrier density.

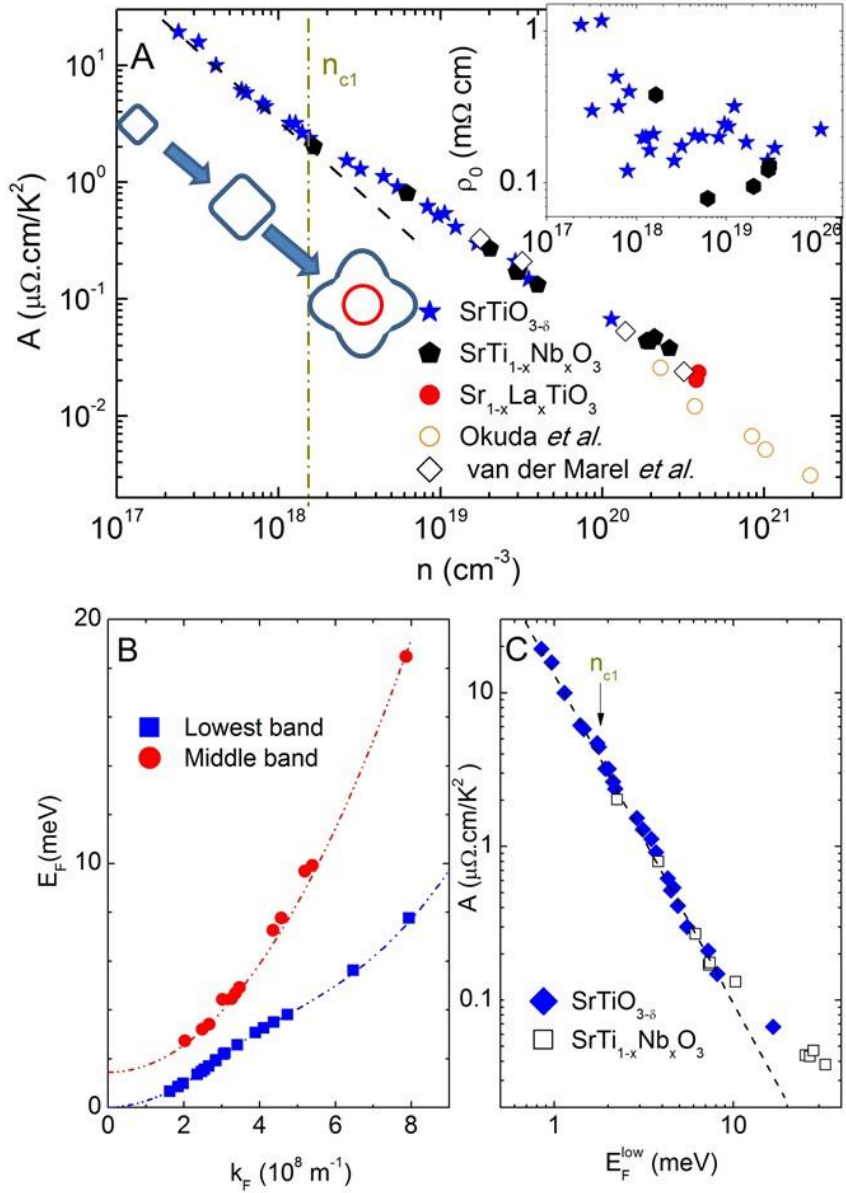

**Fig. 2. Variation of A with carrier concentration and Fermi energy:** (A) The prefactor A of $T^2$ resistivity as a function of carrier concentration on a log-log scale. The data represented by empty circles and diamonds are from Refs. (12) and (13), respectively. A dash-dot vertical line marks the first critical doping, above which a second band begins to be filled (13, 16). The evolution of Fermi surface with increasing concentration is also sketched. Below $n_{c1}$, the Fermi surface is a simple squeezed ellipsoid. Above, it has two concentric components with the outer growing lobes. Inset: residual resistivity, $\rho_0$ extracted from $\rho=\rho_0+AT^2$ fits. (B) The dispersion of the two bands extracted from quantum-oscillation

measurements (16). (C) The dependence of the prefactor on the Fermi energy measured from the bottom of the lower band. Its dependence is close to $E_F^{-2}$ across $n_{c1}$ with a deviation emerging at higher energies.

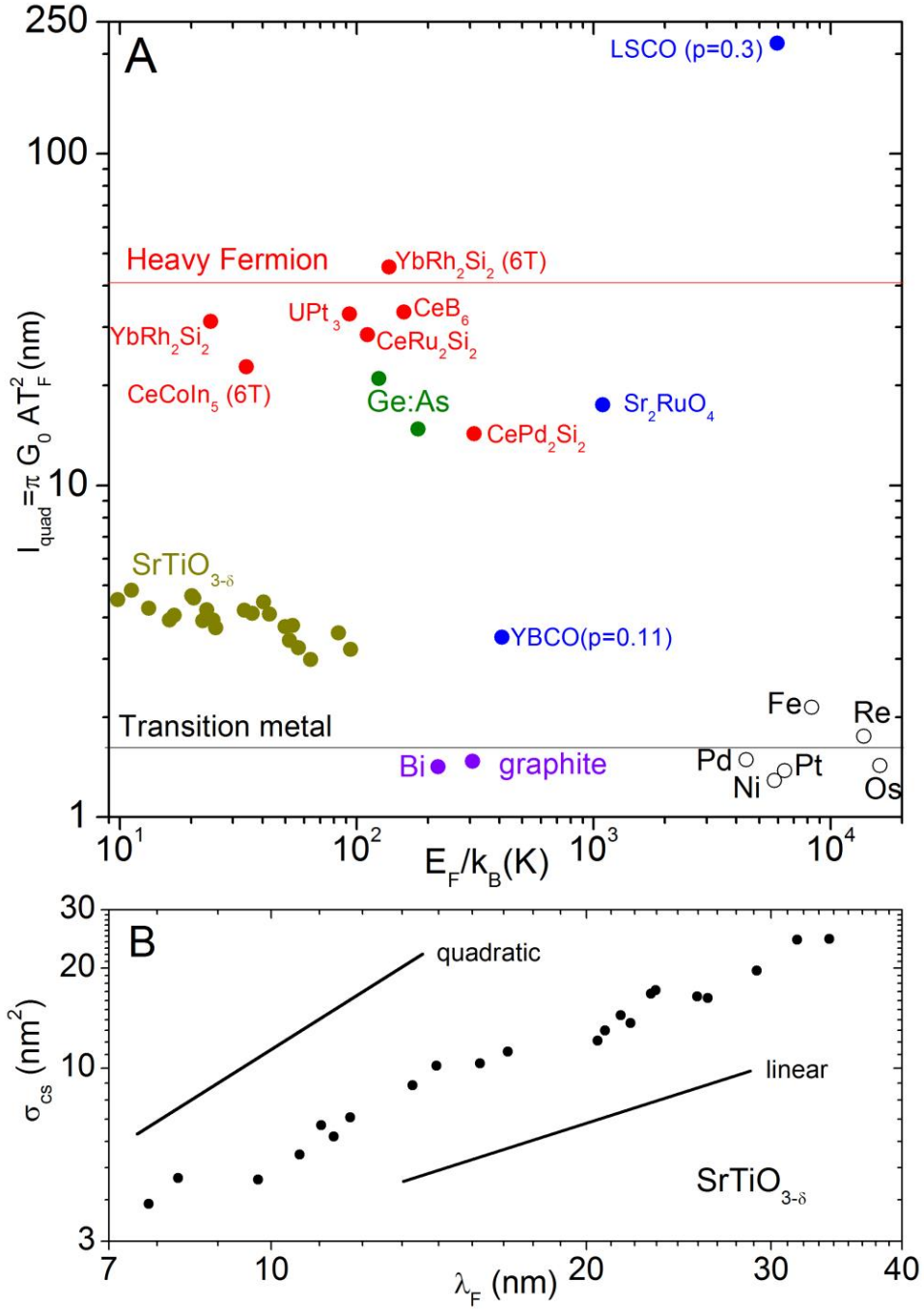

Fig. 3. The characteristic length scale of e-e scattering in $SrTiO_{3-\delta}$ compared to other Fermi liquids: (A) The length scale defined in Eq. 1 and extracted from A and $T_F$ ($l_{quad}=\pi G_0 A T_F^2$ where $G_0=2e^2/h$) in $SrTiO_{3-\delta}$ as well as a number of other Fermi liquids (see Tables S3 and S4 for details and references). The two horizontal solid lines correspond to the Kadowaki-Woods $A/\gamma^2$ ratio in heavy fermions (10 $\mu\Omega cm \cdot mol^2 \cdot K^2 \cdot J^{-2}$ in red) and in transition metals (0.4 $\mu\Omega cm \cdot mol^2 K^2 J^{-2}$ in black) (2, 3, 6).

(B) The extracted collision cross section of electrons (see Eq. 2) as a function of Fermi wave-length follows a dependence close to $\lambda_F^{1.2}$.

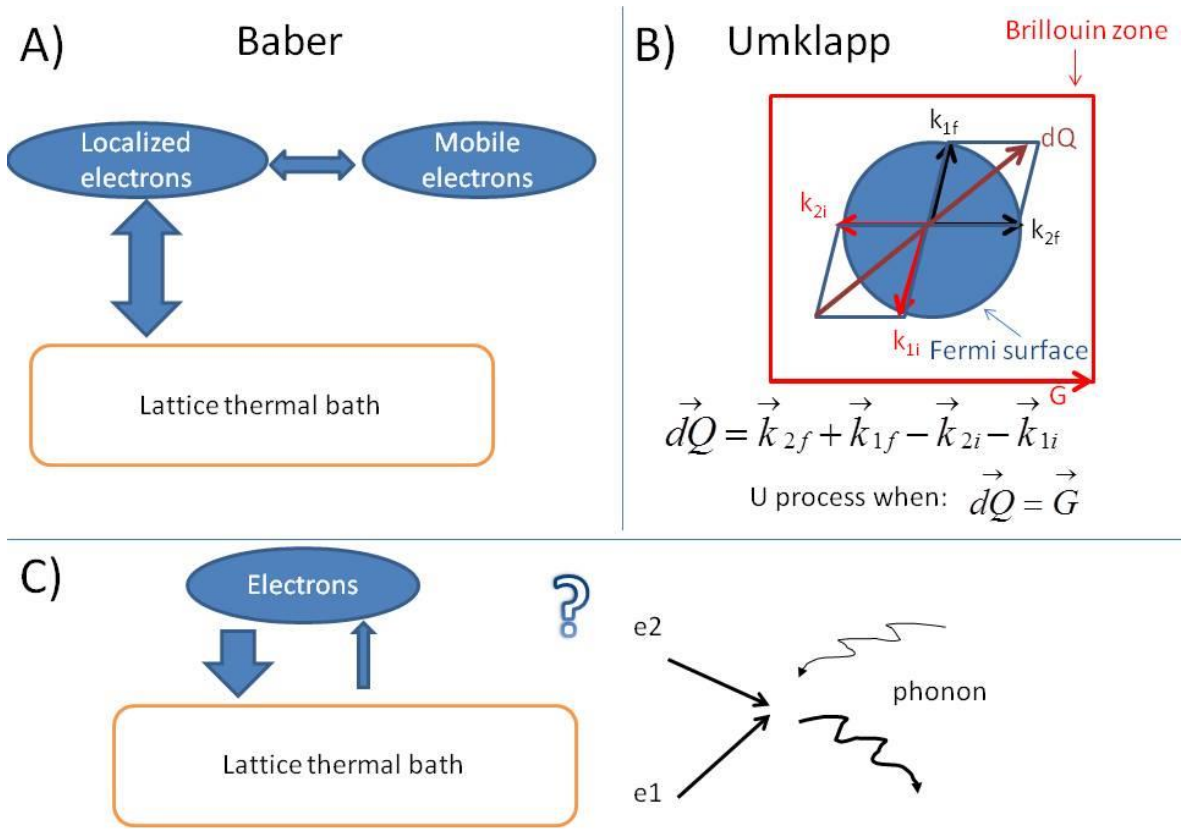

**Fig. 4. Theoretical models for $T^2$ resistivity**: (A) The original mechanism (1) requires two distinct reservoirs of electrons with different strengths of coupling to the lattice. (B) Umklapp scattering in which the momentum balance between incoming and outgoing electrons differs by a unit vector of the reciprocal lattice. Such events are only possible when the Fermi wave-vector is equal or larger than one-fourth of the Brillouin zone width. (C) Neither of these scenarios can explain the persistence of $T^2$ resistivity in the case of dilute $SrTiO_{3-\delta}$, in which there is a single tiny Fermi surface at the center of the Brillouin zone. A scenario is required in which (at least some) scattering events between electrons are accompanied by an asymmetric exchange of momentum with the lattice.

# Supplementary Material
# Scalable T² resistivity in a small single-component Fermi surface

Xiao Lin[1], Benoît Fauqué[1], and Kamran Behnia[1]*

[1]Laboratoire Photons Et Matières (CNRS/UPMC), ESPCI, 10 Rue Vauquelin, F-75005, Paris, France

*Correspondence to: kamran.behnia@espci.fr.

**Materials and Methods:** This study was carried out on bulk commercial $SrTiO_3$, $SrTi_{1-x}Nb_xO_3$ and $Sr_{1-x}La_xTiO_3$ single crystals. In order to introduce oxygen vacancies, $SrTiO_3$ samples were heated in vacuum from 800 C to 1100 C. For carrier concentrations exceeding $10^{19}$ cm$^{-3}$, in order to enhance oxygen deficiency, we included a small titanium disc to the vacuum chamber during the annealing process. Ohmic contacts were made by evaporating gold and heating. Resistivity was measured with a standard four-probe method in the Quantum Design Physical Property Measurement System (PPMS) from 2K to 300K. Quantum-oscillation measurements were performed in a dilution refrigerator inserted inside a 17 T superconducting magnet and were presented in ref. 16. Table S1 lists the various properties of the samples used in the present study.

**T-square resistivity in Nb-doped and La-doped samples:** The data on the Nb-doped and La-doped samples are presented in Fig. S1. Their resistivity followed a $T^2$ behavior with a prefactor comparable to reduced STO at the same concentration.

**Extracting energy dispersion from quantum oscillations:** Table S2 summarizes the data from quantum oscillations in n-doped $SrTiO_3$ (16), which allowed to construct the energy dispersion. According to the Onsager relation, the frequency of quantum oscillations sets the cross section of the Fermi surface at a given doping level. Assuming a circular cross section, this leads to the radius of the Fermi surface at a given doping level, $k_F(n)$. Using the cyclotron mass extracted from the temperature attenuation of oscillations, one can extract the Fermi velocity at a given carrier density: $v_F(n) = \hbar k_F(n)/m^*(n)$. Fermi energy at a given carrier density $E_F(n)$ by integrating the $v_F(k_F)$ curve, which led to Figure 2B.

**Fermi energy and T² resistivity in other systems:** Fig.S2 compares the variation of *A* with Fermi energy in $SrTiO_{3-\delta}$ with other Fermi liquids in which a quadratic resistivity has been observed and the structure of the Fermi surface is experimentally known. The list includes two heavy-electron metals [$UPt_3$ (31, 32) and $CeRu_2Si_2$ (33,34) in which all components of the Fermi surface have been observed], two oxides [$Sr_2RuO_4$ (22, 35) and $La_{1.7}Sr_{0.3}CuO_4$ (23)] and three semi-metals [bismuth(36, 37), $Bi_{0.96}Sb_{0.04}$(38-40) and graphite(41-43) and]. Table S3 details the reported value of frequency of quantum oscillations and cyclotron mass used to extract the Fermi energy using $E_F= (\hbar k_F)^2/2m^*$. In the case of heavily-doped $La_{1.7}Sr_{0.3}CuO_4$, the Fermi energy was estimated from electronic specific heat coefficient ($\gamma$=6.9 mJ/mol/K$^2$) according to Eq.1. In the case of germanium, it was estimated using the effective mass and carrier density.

**Kadowaki-Woods ratio in other systems:** Table S4 details the electronic specific heat ($\gamma$) and the prefactor of $T^2$ resistivity (A) in heavy-fermion and transition metal systems shown in Fig. 3A.

| Samples | $n_H$ cm$^{-3}$ | $E_{F1\text{-estimated}}$ meV | A µΩ.cm/K$^2$ | $\rho_{300K}$ Ω.cm | $\rho_{2K}$ Ω.cm | RRR | $\mu_{H-2K}$ cm$^2$/V/s |
|---|---|---|---|---|---|---|---|
| SrTiO$_{3-\delta}$ | 2.4E+17 | 0.85 | 19.3 | 5.87 | 2.74E-03 | 2142 | 9504 |
| | 3.2E+17 | 0.97 | 15.8 | 5.75 | 1.66E-03 | 3464 | 11766 |
| | 4.1E+17 | 1.14 | 10 | 2.54 | 1.21E-03 | 2099 | 12598 |
| | 5.9E+17 | 1.40 | 6.13 | 1.86 | 1.28E-03 | 1453 | 8276 |
| | 6.3E+17 | 1.46 | 5.8 | 1.84 | 8.22E-04 | 2238 | 12069 |
| | 7.9E+17 | 1.74 | 4.7 | 1.65 | 5.76E-04 | 2865 | 13632 |
| | 8.3E+17 | 1.77 | 4.45 | 1.42 | 5.68E-04 | 2500 | 13289 |
| | 1.17E+18 | 1.93 | 3.2 | 1.1 | 5.01E-04 | 2196 | 10662 |
| | 1.27E+18 | 2.01 | 3.19 | 1.05 | 4.41E-04 | 2381 | 11159 |
| | 1.4E+18 | 2.13 | 2.64 | 0.868 | 4.07E-04 | 2133 | 10969 |
| | 1.55E+18 | 2.19 | 2.37 | 0.837 | 5.43E-04 | 1541 | 7426 |
| | 2.6E+18 | 2.89 | 1.53 | 0.49 | 2.59E-04 | 1892 | 9141 |
| | 3.2E+18 | 3.12 | 1.29 | 0.402 | 2.53E-04 | 1589 | 7720 |
| | 4.5E+18 | 3.48 | 1.12 | 0.326 | 2.22E-04 | 1468 | 6312 |
| | 5.4E+18 | 3.69 | 0.915 | 0.258 | 2.18E-04 | 1183 | 5280 |
| | 8.3E+18 | 4.30 | 0.62 | 0.171 | 1.89E-04 | 905 | 3984 |
| | 9.6E+18 | 4.48 | 0.518 | 0.141 | 4.31E-04 | 327 | 1511 |
| | 1.06E+19 | 4.62 | 0.54 | 0.132 | 2.30E-04 | 574 | 2564 |
| | 1.24E+19 | 4.90 | 0.412 | 0.117 | 3.05E-04 | 384 | 1653 |
| | 1.7E+19 | 5.51 | 0.3 | 0.0775 | 1.98E-04 | 391 | 1879 |
| | 2.9E+19 | 7.23 | 0.21 | 0.0391 | 1.66E-04 | 236 | 1298 |
| | 3.5E+19 | 8.13 | 0.148 | 0.0275 | 1.75E-04 | 157 | 1020 |
| | 1.14E+20 | 16.68 | 0.067 | 0.0109 | 2.48E-04 | 44 | 221 |
| SrTi$_{1-x}$Nb$_x$O$_3$ | 1.6E+18 | 2.24 | 2.02 | 0.964 | 1.00E-04 | 9640 | 37879 |
| | 6.2E+18 | 3.81 | 0.8 | 0.2 | 8.38E-05 | 2387 | 11991 |
| | 2.02E+19 | 6.13 | 0.27 | 0.0616 | 7.93E-05 | 777 | 3902 |
| | 2.96E+19 | 7.29 | 0.17 | 0.0385 | 1.28E-04 | 301 | 1650 |
| | 3.05E+19 | 7.39 | 0.175 | 0.0405 | 1.15E-04 | 352 | 1782 |
| | 3.98E+19 | 10.28 | 0.132 | 0.0252 | 4.90E-05 | 514 | 3205 |
| | 1.88E+20 | 25.25 | 0.0438 | 6.07E-03 | 5.30E-05 | 115 | 627 |
| | 1.95E+20 | 26.70 | 0.043 | 5.99E-03 | 5.10E-05 | 117 | 628 |
| | 2.1E+20 | 27.98 | 0.047 | 6.60E-03 | 7.10E-05 | 93 | 419 |
| | 2.6E+20 | 32.62 | 0.038 | 5.17E-03 | 1.09E-04 | 47 | 221 |
| Sr$_{1-x}$La$_x$TiO$_3$ | 3.81E+20 | 43.17 | 0.0203 | 2.76E-03 | 1.34E-04 | 21 | 122 |
| | 3.93E+20 | 44.33 | 0.0235 | 3.13E-03 | 1.21E-04 | 26 | 131 |

**Table S1**. *Hall carrier concentration ($n_H$), Fermi energy ($E_F$), the prefactor of $T^2$ resistivity (A), resistivity at room temperature ($\rho_{300K}$) and at 2K ($\rho_{2K}$), the ratio of $\rho_{300K}$ to $\rho_{2K}$ (RRR) and the low-temperature electron mobility for samples shown in Fig. 1 and Fig. 2. For each sample, $E_F$ is estimated from the experimentally-resolved energy dispersion and the carrier density.*

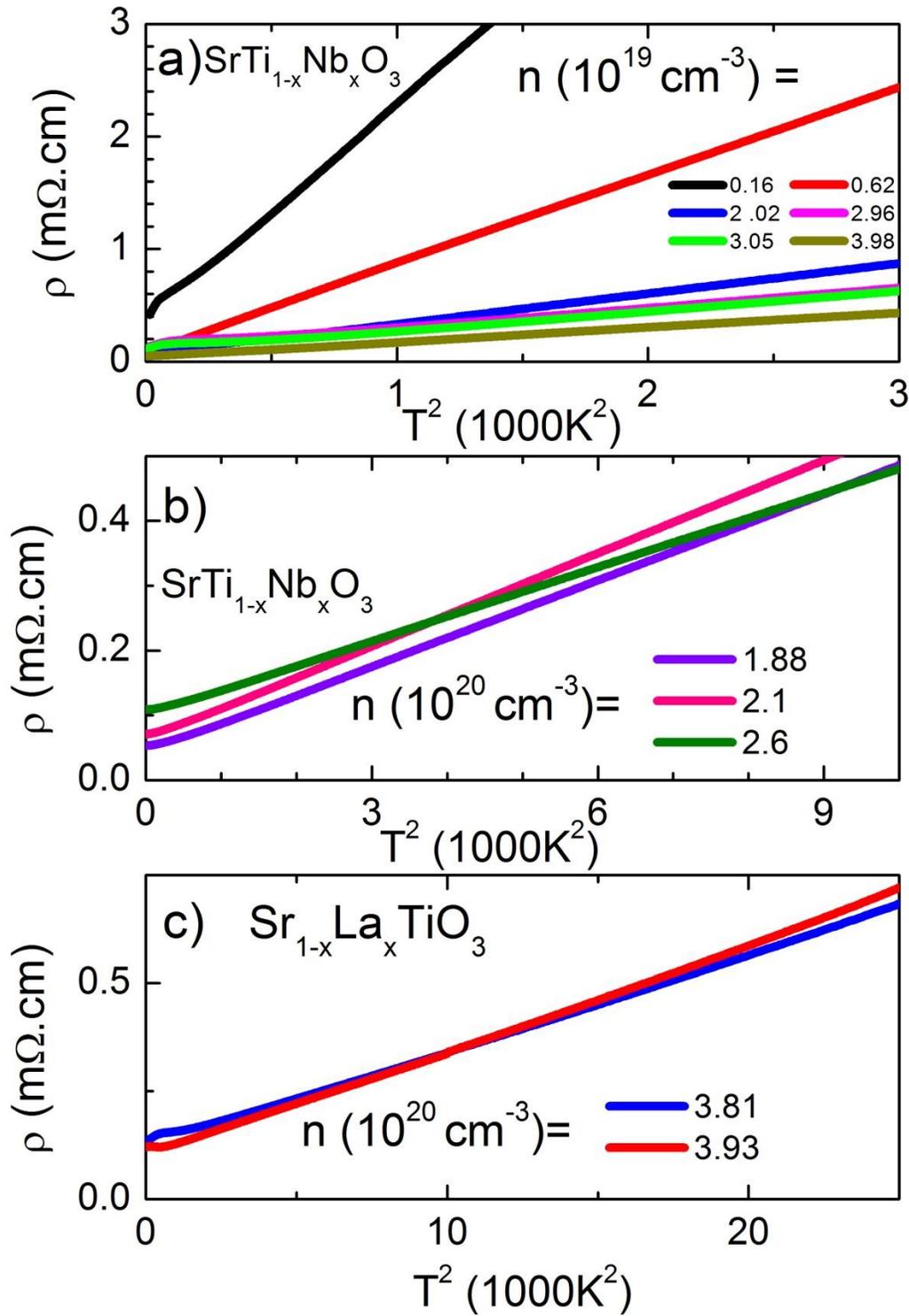

**Fig. S1**. *T-square resistivity and smooth variation of its slope) in Nb-doped (A,B,) and La-doped (C) samples SrTiO₃. Carrier concentration was determined by measuring the Hall coefficient, which was found to be independent of temperature and linear in magnetic field.*

| Samples | n (cm⁻³) | $F_1$ (T) | $m^*_1$ ($m_e$) | $E_{F1}$ (meV) | $F_2$ (T) | $m^*_2$ ($m_e$) | $E_{F2}+\Delta_{12}$ (meV) | $F_3$ (T) | $m^*_3$ ($m_e$) |
|---|---|---|---|---|---|---|---|---|---|
| SrTiO$_{3-\delta}$ | 1.58E+17 | 8.7 | 1.5(0.1) | 0.672 | - | - | - | - | - |
| | 2.40E+17 | 11.2 | 1.5(0.1) | 0.866 | - | - | - | - | - |
| | 3.20E+17 | 12.9 | 1.5(0.05) | 0.997 | - | - | - | - | - |
| | 5.50E+17 | 18.2 | 1.83(0.07) | 1.37 | - | - | - | - | - |
| | 6.30E+17 | 20 | 1.7(0.1) | 1.49 | - | - | - | - | - |
| | 6.79E+17 | 21.3 | 1.74(0.1) | 1.57 | - | - | - | - | - |
| | 7.43E+17 | 23.25 | 1.7(0.2) | 1.70 | - | - | - | - | - |
| | 1.06E+18 | 26.5 | 1.7(0.3) | 1.93 | - | - | - | - | - |
| | 1.20E+18 | 26.8 | 1.8(0.3) | 1.95 | - | - | - | - | - |
| | 1.46E+18 | 30.7 | 1.9(0.2) | 2.19 | - | - | - | - | - |
| | 1.65E+18 | 31.4 | 2.3(0.4) | 2.23 | 7.1 | N.D. | N.D. | - | - |
| | 1.93E+18 | 38.4 | 2.4(0.25) | 2.57 | 13.6 | 1.7(0.25) | 2.73 | - | - |
| | 2.88E+18 | 49.8 | 3.48(0.15) | 3.07 | 20.4 | 1.64(0.09) | 3.20 | - | - |
| | 3.88E+18 | 55.5 | 3.3(0.2) | 3.27 | 23.4 | 1.65(0.15) | 3.41 | - | - |
| | 4.11E+18 | 63 | 4(0.5) | 3.51 | 28 | N.D. | N.D. | - | - |
| | 7.66E+18 | 84 | N.D | N.D | 35 | 1.09(0.05) | 4.46 | - | - |
| | 8.30E+18 | N.D | N.D | N.D | 37.2 | 1.15(0.07) | 4.69 | - | - |
| | 9.60E+18 | N.D | N.D | N.D | 39.6 | 1.2(0.2) | 4.92 | - | - |
| | 1.68E+19 | N.D | N.D | N.D | 62.5 | 1.08(0.15) | 7.27 | - | - |
| | 2.90E+19 | N.D | N.D | N.D | 89 | 1.5(0.1) | 9.68 | - | - |
| SrTi$_{1-x}$Nb$_x$O$_3$ | 6.22E+18 | 74 | 3(0.1) | 3.81 | 30 | 1.06(0.1) | 4.43 | - | - |
| | 1.60E+19 | 138 | 4.2(0.2) | 5.62 | 69 | 1.65(0.25) | 7.77 | - | - |
| | 3.18E+19 | 208 | 3.5(0.13) | 7.77 | 96 | 1.33(0.4) | 9.92 | 7 | N.D |
| | 1.60E+20 | N.D. | N.D. | N.D. | 204 | 1.57(0.08) | 18.48 | 116 | 1.55(0.08) |

**Table S2.** *Hall carrier concentration, quantum oscillation frequency, cyclotron mass and Fermi energy for samples giving rise to Fig. 2b. In this table, N.D. refers to quantities which could not be determined. $\Delta_{12}$ refers to the energy difference between the bottom of middle band and lowest band.*

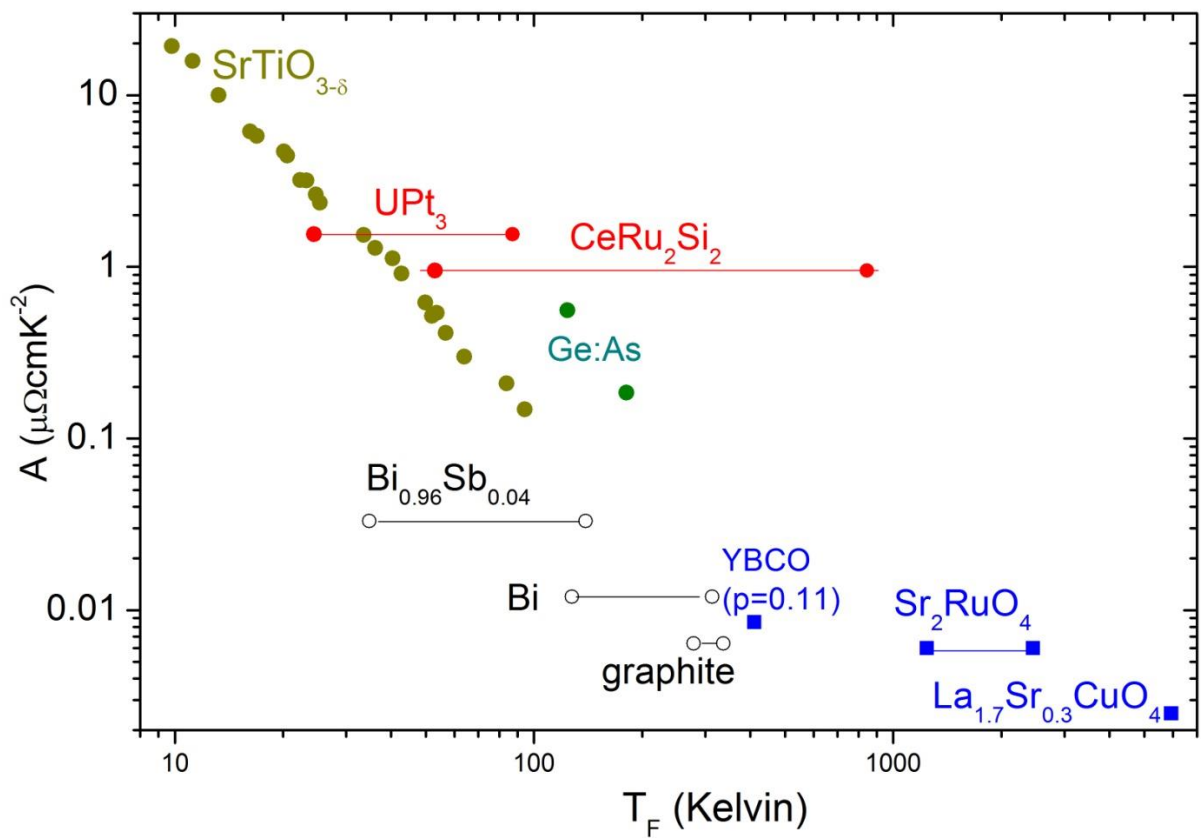

**Fig. S2**. The magnitude of *A* vs. Fermi energy in a number of Fermi liquids. The details are given in Table S3.

| System | Bands | | Frequency T | m* $m_e$ | $E_F/k_B$ K | A $\mu\Omega.cm/K^2$ |
|---|---|---|---|---|---|---|
| UPt₃ (31, 32) | α | | 540 | 25 | 29 | 10.55 |
| | γ | | 730 | 40 | 24 | |
| | δ | | 1400 | 50 | 37 | |
| | ε | | 2100 | 60 | 47 | |
| | ω | | 5850 | 90 | 87 | |
| CeRu₂Si₂ (33, 34) | 001 | β | 536 | 1.5 | 480 | 0.94 |
| | | γ | 980 | 1.6 | 824 | |
| | | κ | 1650 | 11 | 202 | |
| | | μ | 2.69E4 | 50 | 723 | |
| | 110 | α | 484 | 12.3 | 53 | |
| | | β | 970 | 1.8 | 725 | |
| | | γ | 1420 | 2.3 | 830 | |
| | | ε | 2540 | 20 | 171 | |
| | | ψ* | 8000 | 140 | 76.8 | |
| | 100 | α | 452 | 15 | 40 | |
| | | β | 943 | 1.5 | 845 | |
| | | γ | 1570 | 2.8 | 754 | |
| | | ψ* | 4840 | 120 | 54 | |
| | | ψ | 5420 | 120 | 61 | |
| Sr₂RuO₄ (22,35) | α-h | | 3047 | 3.3 | 1240 | 6.1E-3 |
| | β-e | | 12755 | 7 | 2450 | |
| | γ-e | | 18693 | 16 | 1571 | |
| Bi (27,36,37) | Binary | e | 18.9 | 0.0272 | h 127.6 | 0.012 |
| | | e | 1.4 | 0.0021 | | |
| | | h | 22.2 | 0.221 | | |
| | Bisectrix | e | 1.2 | 0.0018 | | |
| | | e | 2.4 | 0.0037 | | |
| | | h | 22.2 | 0.221 | e 313.2 | |
| | Trigonal | e | 8.5 | 0.0125 | | |
| | | h | 6.3 | 0.0678 | | |
| Bi₀.₉₆Sb₀.₀₄ (38-40) | Binary | h | 3.33 | -- | h 34.8 | 0.033 |
| | Bisectrix | e | 0.32 | 5E-4 | | |
| | | h | 3.33 | -- | | |
| | Trigonal | e | 1.8 | 0.0032 | e 140.4 | |
| | | h | 2 | -- | | |
| Graphite (41-43) | e | | 6.15 | 0.054 | 278 | 6.4E-3 |
| | h | | 4.5 | 0.039 | 336 | |
| YBCO(p=0.11) (24-26) | | | 540 | 1.76 | 410 | 8.5E-3 |
| Ge:As (29) | | | -- | -- | 181;124 | 0.185;0.56 |
| La₁.₇Sr₀.₃CuO₄ (23) | | | -- | -- | 7700 | 2.5E-3 |

**Table S3.** *Reported frequencies and effective masses, Fermi energy deuced and the prefactor of $T^2$ resistivity for systems shown in Fig. S2. Electron and hole pockets are designated by e and h and different Fermi surfaces are marked by (α, β…). Fermi energy was calculated using $E_F = (\hbar k_F)^2/2m^*$ with $k_F$ estimated from frequencies. In CeRu₂Si₂, [100], [001] refer to different orientations.*

| System | γ (mJ/mol/ K$^2$) | A (μΩ.cm/K$^2$) |
|---|---|---|
| CePd$_2$Si$_2$ (44, 45) | 131 | 0.06 |
| CeB$_6$ (46, 47) | 250 | 0.832 |
| YbRh$_2$Si$_2$ (6T) (48) | 300 | 1.0 |
| CeRu$_2$Si$_2$ (33, 49) | 354 | 0.94 |
| UPt$_3$ (31, 32) | 440 | 1.55 |
| CeCoIn$_5$ (6T) (50, 51) | 1450 | 7.6 |
| YbRh$_2$Si$_2$ (52) | 1700 | 22 |
| Os (2, 53) | 2.5 | 2.2E-6 |
| Re (53, 54) | 3.0 | 3.8E-6 |
| Fe (2,55) | 4.9 | 1.3E-5 |
| Pt (53,55,56) | 6.4 | 1.4E-5 to 2E-5 |
| Ni (53,57,58) | 7.1 | 9.5E-6 to 2.6E-5 |
| Pd (53,59) | 9.3 | 3.3E-5 |

**Table S4.** *Electronic specific heat (γ) and the prefactor of T$^2$ resistivity (A) of heavy Fermion and transition metal systems presented in Fig. 3B.*